\documentclass[letter,scriptaddress,twocolumn,prl,showpacs,showkeys]{revtex4-1}
		\usepackage{epsfig}

\begin{document}

\preprint{}

\title{Constraining Baryon--Dark Matter Scattering with the Cosmic Dawn 21-cm Signal}

\author{Anastasia Fialkov}
\affiliation{Harvard-Smithsonian Center for Astrophysics, 60 Garden Street, Cambridge, MA 02138, USA}
\author{Rennan Barkana}
\author{Aviad Cohen}
\affiliation{Raymond and Beverly Sackler School of Physics and Astronomy, Tel Aviv University, Tel Aviv 69978, Israel}

\date{\today}

\begin{abstract}
The recent detection of an anomalously strong 21-cm signal of neutral
hydrogen from Cosmic Dawn by the EDGES Low-Band radio experiment can
be explained if cold dark matter particles scattered off the baryons
draining excess energy from the gas. In this Letter we explore the
expanded range of the 21-cm signal that is opened up by this
interaction, varying the astrophysical parameters as well as the
properties of dark matter particles in the widest possible range. We
identify models consistent with current data by comparing to both the
detection in the Low-Band and the upper limits from the EDGES
High-Band antenna. We find that consistent models predict a 21-cm
fluctuation during Cosmic Dawn that is between 3 and 30 times larger
than the largest previously expected without dark matter
scattering. The expected power spectrum exhibits strong Baryon
Acoustic Oscillations imprinted by the velocity-dependent
cross-section. The latter signature is a smoking gun of the
velocity-dependent scattering and could be used by interferometers to
verify the dark matter explanation of the EDGES detection.

\end{abstract}


\maketitle {\it Introduction:} The first few hundred million years
after the Big Bang are the least explored period in the history of the
Universe. This epoch is marked by some of the most interesting events
in cosmic history such as the formation of the very first stars and
black holes. However, what makes this epoch even more attractive for
observers and theorists alike is that dark matter might manifest
itself differently than today in the unmatched physical conditions of
the early Universe. Even though conventional dark matter models assume
only gravitational interactions with ordinary baryonic matter, other
forms of couplings, such as collisions between dark matter and gas,
are not ruled out and could be important at early times when the
density of matter was much higher than today. Such interactions could
modify the thermal and ionization histories, leaving fingerprints in
the 21-cm signal of atomic hydrogen \cite{Tashiro:2014, Munoz:2015,
  Barkana:2018}.

In the standard cosmology, Compton scattering couples the baryon
temperature, $T_{\rm gas}$, to the temperature of the Cosmic Microwave
Background (CMB) at redshifts above $z_{\rm dec}\sim 200$. At lower
redshifts, $T_{\rm gas}$ is expected to cool adiabatically due to the
expansion of the Universe until the moment when the first X-ray
sources turn on, injecting energy into the gas. Because dark matter is
expected to decouple earlier and be much colder than the gas,
collisions between baryons and dark matter particles could drain
excess energy from the gas leading to its over-cooling
\cite{Tashiro:2014}, while the relative velocity between dark matter
and the gas could in some cases result in overall over-heating of the
baryons \cite{Munoz:2015}.

The 21-cm line of neutral atomic hydrogen with a rest-frame frequency
of 1.42 GHz is one of the most promising probes of this epoch. This
signal is sensitive to the thermal and ionization states of the
baryons and, thus, can be used to measure the energy balance of the
early Universe. The brightness temperature of the 21-cm line,
$T_{21}$, is coupled to the kinetic temperature of the baryons by two
processes: During the cosmic Dark Ages ($35\lesssim z\lesssim 1100$,
the epoch preceding the formation of the very first stars) the gas is
dense enough for interatomic collisions to drive the effective
temperature of the 21-cm transition to the temperature of the gas, a
process that becomes less efficient as the Universe expands. During
the subsequent epoch of Cosmic Dawn ($15\lesssim z\lesssim 35$) when
the first stars are formed, the main driver is the Ly-$\alpha$
radiation produced by stars which couples the two temperatures via the
Wouthuysen-Field process \cite{Wouthuysen:1952, Field:1958}. Because
gas is colder than the background CMB at $z<z_{\rm dec}$ and before the
first X-ray sources turn on, the 21-cm signal from the Dark Ages and
Cosmic Dawn is seen in absorption against the CMB. Features of the
high-redshift 21-cm signal depend on the underlying astrophysics
\cite{Cohen:2017a}, but also on the properties of dark matter
particles, if the latter affect either the thermal or the ionization
state of the gas \cite{Tashiro:2014, Munoz:2015,
  Barkana:2018}. Therefore, the 21-cm signal can be used as a unique
probe of the dark sector.

Although exploration of the high-redshift domain is one of the most
active areas of research in astrophysics, properties of the early
Universe are still poorly constrained. The uncertainty in
astrophysical parameters and the limited understanding of the dark
matter physics propagate into the 21-cm modeling and result in a large
variety of allowed signals. The dependence of the expected 21-cm
signal on astrophysical parameters has been extensively explored
\cite{Cohen:2017a, Cohen:2017b, Cohen:2018}. It is the goal of this
Letter to explore the 21-cm signal over the parameter space of both
astro- and dark matter physics.

{\it Observation and Theory:} The first detection of the 21-cm signal
from Cosmic Dawn was recently reported by the Law-Band antenna of the
Experiment to Detect the Global EoR Signature (EDGES)
\cite{Bowman:2018}. After removal of the foregrounds and the
instrumental noise, excess signal was found in the data of the
Low-Band antenna observing in the 50-100 MHz frequency band and
probing the 21-cm signal from the redshift range of $z\approx
13.2-27.4$. The extracted cosmological signal, centered at $\nu =
78\pm 2$ MHz (which corresponds to $z = 17.2$), features a broad
absorption trough of $T_{21} = -500^{+200}_{-500}$ mK, where the error
corresponds to $99\%$ confidence including both thermal and systematic
noise.  In the standard cosmological scenario, the strongest possible
absorption expected at $z\sim 17$ is $-209$ mK, which corresponds to a
gas temperature of $T_{\rm gas}\sim 7$ K. The observed $T_{21}<300$ mK
requires the gas to be much colder, $T_{\rm gas}<5.1$ K, 
which is hard to explain by astrophysics alone \cite{Barkana:2018}. 
Feng \& Holder \cite{Feng:2018} suggest that an excess radio background, 
such as seen by ARCADE 2 \cite{Fixsen:2011}, could produce anomalously 
strong absorption in the 21-cm signal at $z\sim 20$. However, the ARCADE 2 excess  
alone does not require astrophysical  explanation and can be explained
 by carefully modeling the galactic contribution
\cite{Subrahmanyan:2013}.

To explain the anomalously strong absorption seen by EDGES Low,
Barkana (2018) \cite{Barkana:2018} invoked elastic velocity-dependent
scattering between baryons and dark matter (b-DM scattering). The
absorption trough as deep as detected by EDGES is obtained if baryons
scatter off dark matter particles with masses in the range $m_\chi
<4.3$ GeV. The scattering cross-section is assumed to be
velocity-dependent, $\sigma(v)=
\sigma_1\left(v/1\textrm{km~s}^{-1}\right)^{-n}$, where $\sigma_1>
3.4\times 10^{-21}$ cm$^2$ is the normalization assuming $n=4$, and
$v$ is the relative velocity between the baryon and the dark matter
particle. Even though the results are derived for $n=4$ (a theory
often considered in the literature and corresponding to a Coulomb-like
b-DM scattering) the qualitative conclusion is more general and is not
limited by the specific type of scattering. Astrophysics also plays an
important role, as the deep absorption in the EDGES Low-Band is
produced only in the presence of the stellar Ly-$\alpha$ background;
the central frequency and the depth of the trough are determined, in
addition to b-DM scattering, by the timing and intensity of both the
Ly-$\alpha$ and the X-ray radiative backgrounds.

A major role is played by the remnant b-DM relative velocity $v_{\rm bdm}$
from the early universe \cite{Munoz:2015, Barkana:2018}. As a result
of the velocity-dependent scattering, $T_{21}$ is expected to be
modulated by the velocity field: regions where the velocity is low
cool stronger and exhibit a stronger absorption signal; while regions
where the velocity is high cool less. Examples of this dependence are
shown in Fig.~1 and the effect is discussed further in the text.

\begin{figure*}
\begin{center}
\includegraphics[width=3in]{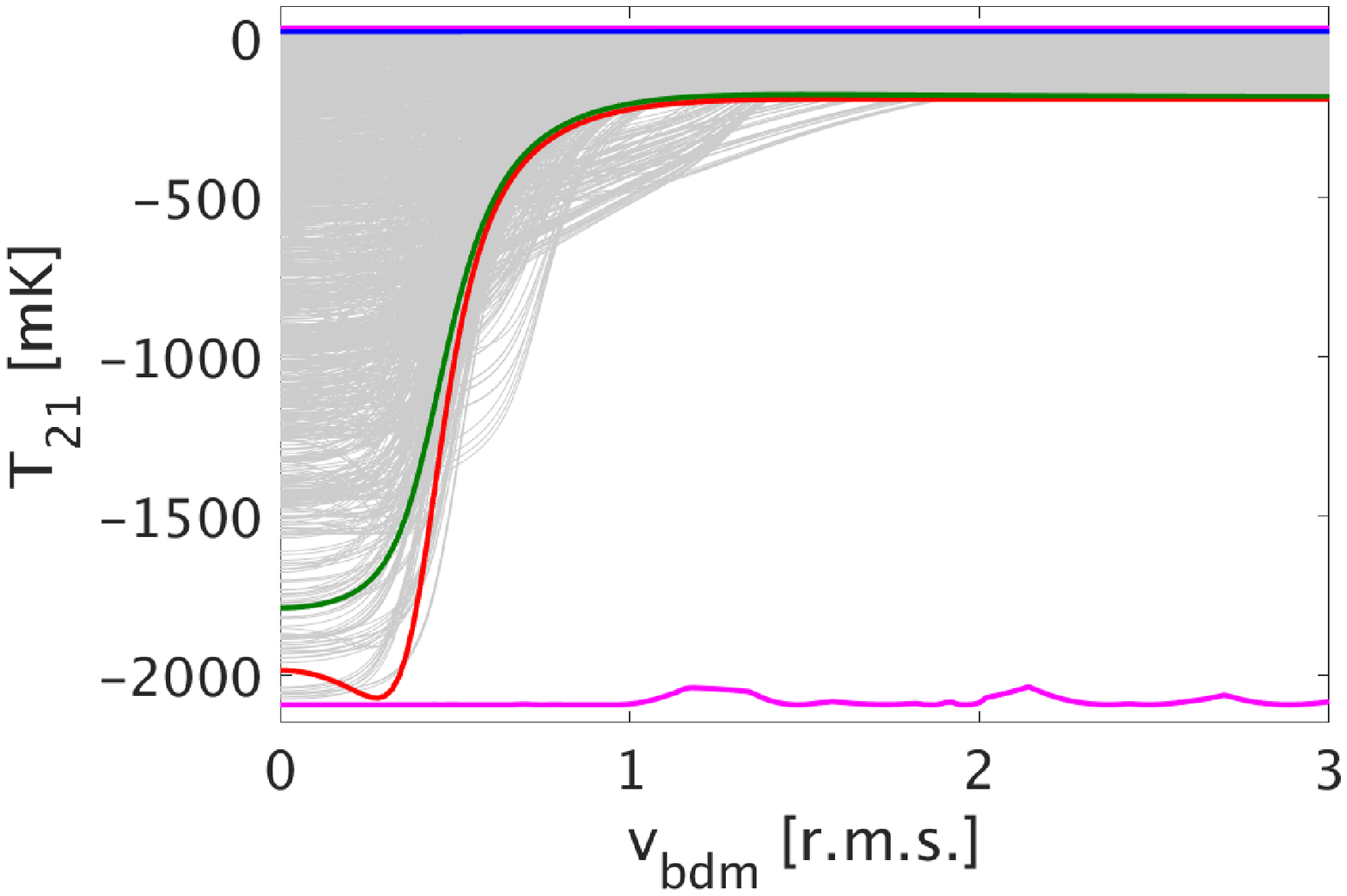}\includegraphics[width=3in]{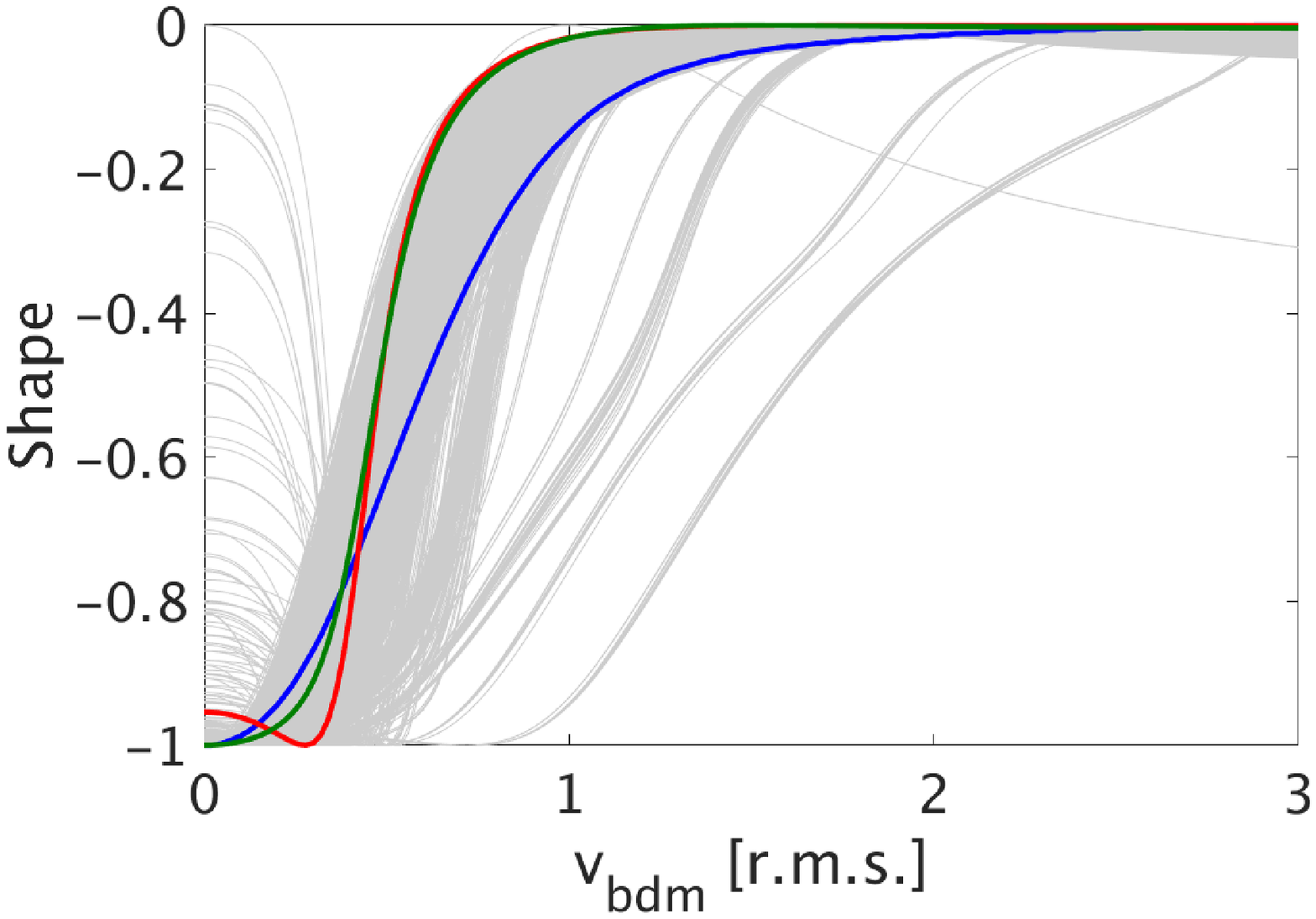}
\includegraphics[width=3in]{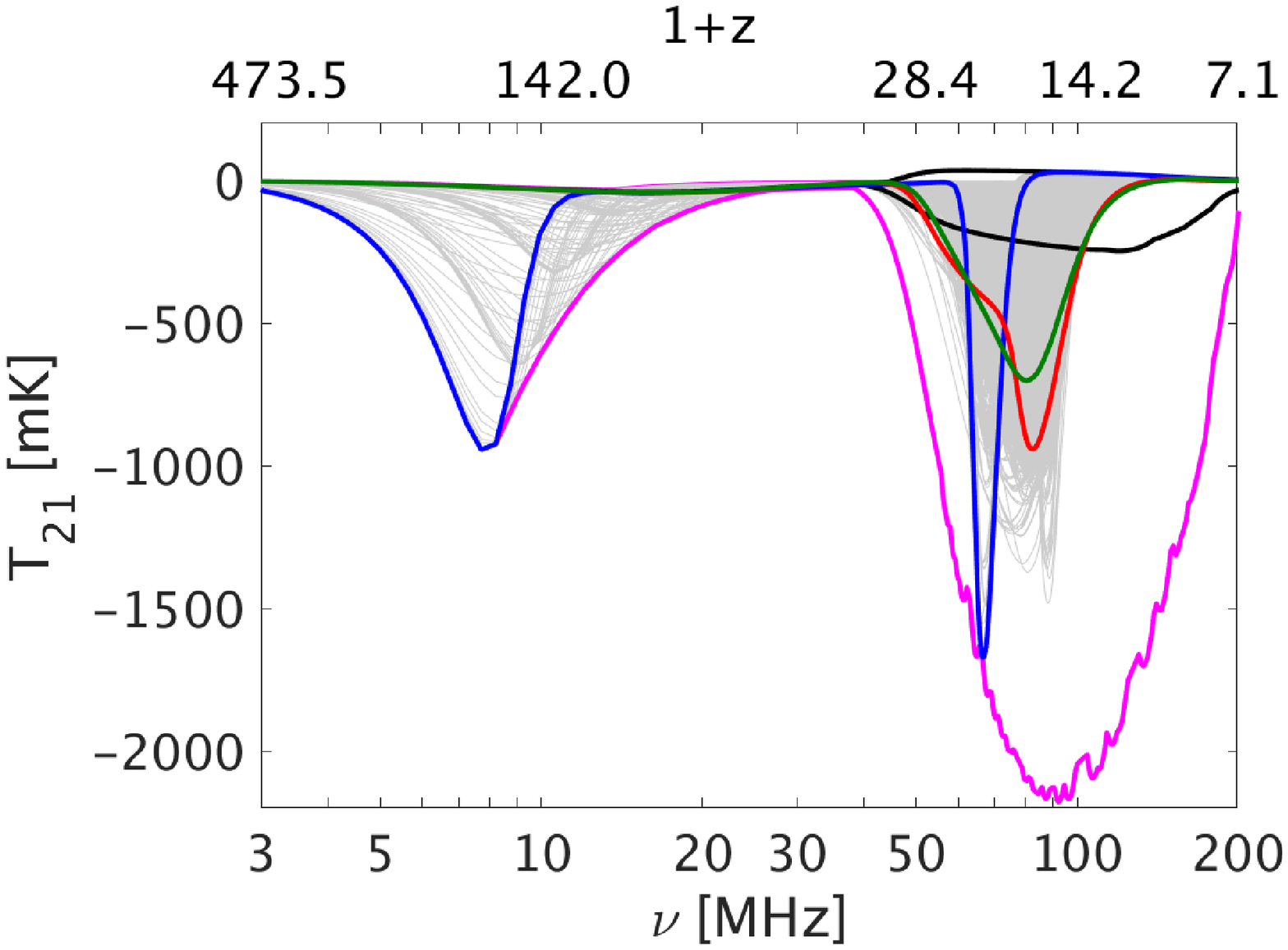}\includegraphics[width=3in]{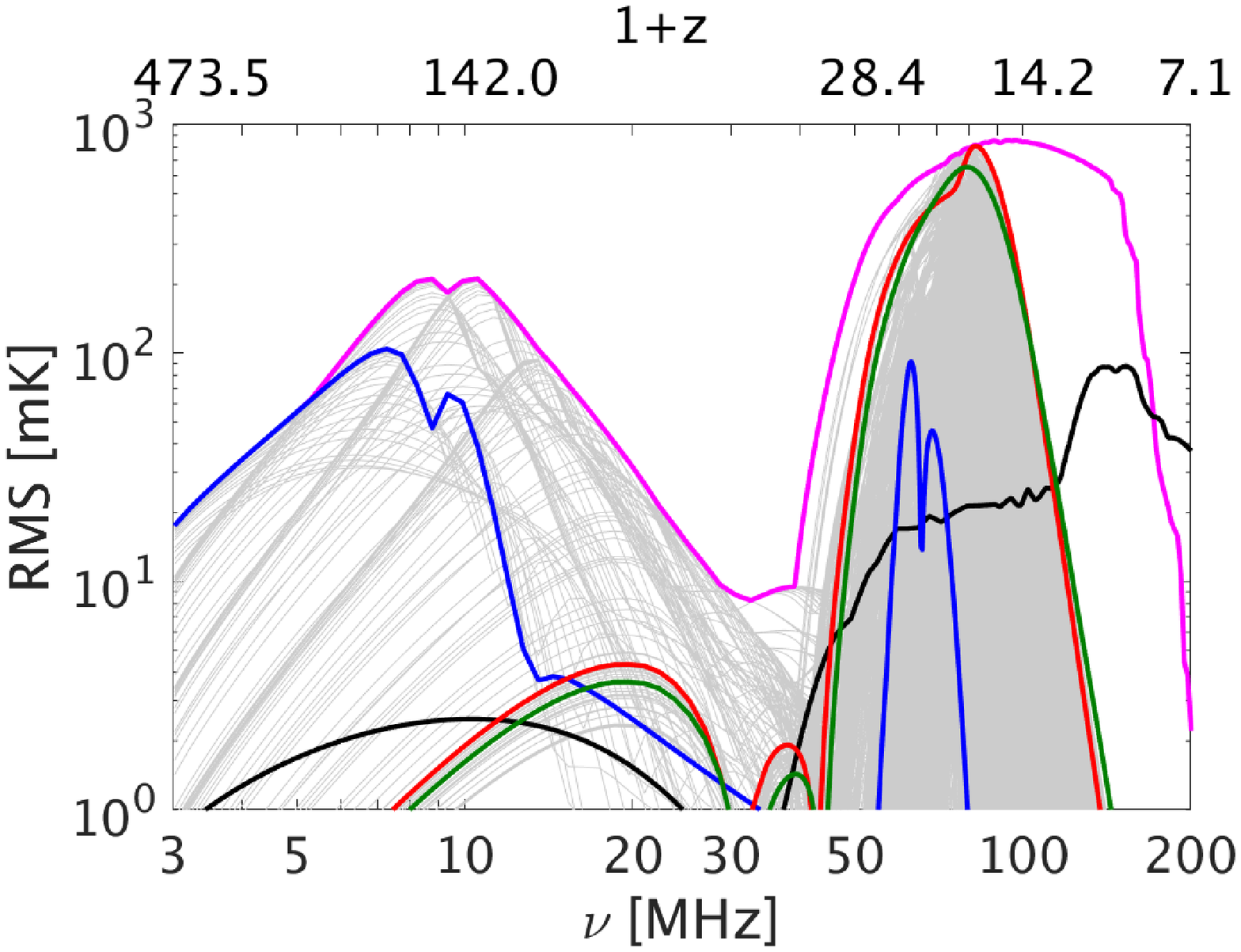}
\caption{{\bf Top:} The function $T_{21}(v_{\rm bdm})$ at $z=17$ in units
  of mK (left) and its shape (right) versus the local
  relative baryon-DM velocity $v_{\rm bdm}$ in units of the
  r.m.s.\ velocity. To show the shape we re-scaled each curve along
  the $y$-axis so that it ranges from $-1$ to 0. {\bf Bottom:} Global
  (i.e., velocity-averaged) 21-cm signal (left) and r.m.s.\ of its
  fluctuations (right), as a function of frequency (bottom $x$-axis) or
  redshift (top $x$-axis). In every panel we show the envelope of all
  the possible signals without b-dm scattering (black, 6389 models)
  and with the scattering included (magenta, 325839 models). Also
  shown are all the models consistent with EDGES Low-Band and
  High-Band (grey lines, 3046 models in total). Out of the latter
  compilation we highlight three models: the model with the deepest
  absorption trough (blue) which is characterize by (see text) $V_c =
  16.5$ [km s$^{-1}$], $f_*=0.5$\%, $f_X = 10$, $\alpha = -1$,
  $\nu_{\rm min}=1$ [keV], $\tau = 0.0703$, $R_{\rm mfp}=20$ [Mpc],
  $m_\chi=0.0032$ [GeV] and $\sigma_1 = 316\times 10^{-20}$ [cm$^2$];
  the lowest redshift of heating transition (green) with $V_c = 16.5$
  [km s$^{-1}$], $f_*=0.3$\%, $f_X = 0.0721$, $\alpha = -1$,
  $\nu_{\rm min}=1$ [keV], $\tau = 0.0702$, $R_{\rm mfp}=30$ [Mpc], $m_\chi=0.1$
  [GeV] and $\sigma_1 = 1\times 10^{-20}$ [cm$^2$]; and the highest
  r.m.s.\ (red) with $V_c = 16.5$ [km s$^{-1}$], $f_*=0.5$\%, $f_X =
  0.01$, $\alpha = -1$, $\nu_{\rm min}=0.1$ [keV], $\tau = 0.0775$,
  $R_{\rm mfp}=20$ [Mpc], $m_\chi=0.0032$ [GeV] and $\sigma_1 = 1\times
  10^{-20}$ [cm$^2$]. }
\end{center}
\end{figure*}

{\it Cosmic Dawn Signal:} We model the 21-cm signal as in
\cite{Barkana:2018} over the large parameter space of possible
astrophysical and dark matter properties. To describe astrophysics we
use seven free parameters (see ref. \cite{Cohen:2018} for more
details). Star formation is parameterized by two numbers: per-halo
efficiency, $0.05\% \leq f_* \leq 50$\%, and minimum circular velocity
(or, equivalently, the minimal cooling mass) of star forming halos,
$4.2\leq V_c\leq 76.5$ km s$^{-1}$. We use three parameters to
describe the X-ray population. These include the slope and the
low-energy cutoff ($-1.5\leq\alpha\leq -1$, $0.1\leq \nu_{\rm min}\leq 3$
keV) of the spectral energy distribution, which is assumed to have a
power law shape. The total X-ray luminosity is assumed to scale as the
star formation rate and is parameterized by $f_X$ with $10^{-4}\leq
f_X \leq 1000$; $f_X = 1$ corresponds to the luminosity of
low-redshift X-ray counterparts, with the redshift evolution of
metallicity taken into account \cite{Fragos:2013, Fialkov:2014}.  The
ionizing efficiency of sources (calibrated to produce the total CMB
optical depth consistent with the Planck data \cite{Ade:2016}) and the
mean free path of ionizing photons ($20\leq R_{\rm mfp}\leq 40$ comoving
Mpc) are the other two free parameters. For the study presented here
we use a compilation of 6389 different astrophysical models populating
the entire parameter space. For each astrophysical scenario we ran 51
models to include the dark matter physics. The b-DM scattering adds
two parameters: the mass of dark matter particles,
$0.0032<m_{\chi}<100$ GeV, and the cross-section
$10^{-30}<\sigma_1<3.16\times 10^{-18}$ cm$^2$. The range of each of
the parameters is consistent with existing observational limits \cite{Barkana:2018}.

In the absence of scattering, the 21-cm signal is calculated using a
state-of-the-art semi-numerical code \cite{Fialkov:2014,
  Cohen:2017a}. Given a set of astrophysical parameters, the
simulation generates histories of the 21-cm signal in comoving volumes
of 384$^3$ Mpc$^3$ resolved down to 3 comoving Mpc. On smaller scales
sub-grid models are employed. Processes such as the growth of
structure, star formation, heating and ionization are incorporated. In
the calculation, large scale structure is tracked from $z \sim 60$,
just after the first stars are expected to form in the observable
Universe, down to $z\sim 6$ when neutral gas is completely
reionized. The reionization history is calibrated to observations
\cite{Ade:2016}. Stars and stellar remnants produce inhomogeneous
X-ray, Ly-$\alpha$, Lyman-Werner and ultraviolet radiative
backgrounds. The impact of radiation on the environment is calculated,
including heating of the intergalactic medium by X-rays.

The b-DM scattering changes the energy budget of the gas. To calculate
the complete 21-cm signal we follow the approach taken by Barkana
(2018) \cite{Barkana:2018}. We calculate the gas temperature
accounting for both the scattering term (as a function of the local
value of relative velocity $v_{\rm bdm}$) and the astrophysical heating
rate. The 21-cm signal is then calculated using the corresponding
stellar Ly-$\alpha$ and ionizing backgrounds. We calculate the global
21-cm signal (as observed by EDGES) averaging over the value of
$v_{\rm bdm}$ which is drawn from the Maxwell-Boltzmann distribution. In
addition to the globally-averaged signal we estimate the r.m.s.\  of
its fluctuations from the Dark Ages and Cosmic Dawn. For the purpose
of this Letter we neglect the 21-cm fluctuations due to the density
and inhomogeneous astrophysical radiation fields, such as X-ray and
Ly-$\alpha$ backgrounds, due to the much larger fluctuations
induced by the velocity-dependent cross-section.

{\it Results:} Fig.~1 shows the range at redshift 17 of the global
(sky-averaged) signal $T_{21}(z)$ and the r.m.s.\ of the fluctuations
expected from the entirety of the considered models, with and without
the contribution from b-DM scattering. The variation on the sky is
determined by the brightness temperature as a function of the local
$v_{\rm bdm}$; this function is also shown in the figure. First, to
highlight the importance of the scattering process we show an envelope
of the maximal and minimal $T_{21}(z)$ as well as the maximal
r.m.s.\ of the fluctuations for the entire ensemble of the 6389
astrophysical cases without b-DM scattering (these do include the
astrophysical fluctuation sources), and 325839 cases including the
scattering. The recent detection by EDGES Low-Band \cite{Bowman:2018}
and the non-detection by EDGES High-Band \cite{Monsalve:2017}
constrain the space of both astrophysical and dark matter
parameters. Here we only verify a rough agreement with the data by
imposing the following cuts based on published $99\%$ or $3\sigma$
limits: The data collected by EDGES High-Band rules out models with
large variations and implies $|T_{21}(100 MHz)-T_{21}(150
MHz)|<300$~mK \cite{Monsalve:2017}. EDGES Low-Band data give a positive detection and
require the absorption feature to be deep, broad and localized within
the band \cite{Bowman:2018}. Within $99\%$ confidence, the cosmological signal should
satisfy (i) $300~ \rm mK < max \left[T_{21}(62<\nu<68)\right]-
min\left[T_{21}(71<\nu<85)\right] < 1000$ mK, and (ii) $300~ \rm mK <
max\left[T_{21}(92<\nu<98)\right]-min\left[T_{21}(71<\nu<85)\right] <
1000$ mK.  There are 3046 models in total, all shown in Fig.~1 (grey
curves), that satisfy both sets of conditions. Another global signal experiment, 
LEDA \cite{Bernardi:2016}, reported $2\sigma$ limit of $-890$ mK on the 
amplitude of $T_{21}$ at $z\sim 20$, which could also be used to rule out
 extreme cases of baryon over-cooling. However, here we rely only on the EDGES data 
which provides stronger constraints.

We first examine all the considered models with and without b-DM
scattering. Unlike in the cases of negligible scattering in which the
shape of the signal is universal and is described by an absorption
trough followed in some cases by an emission feature
\cite{Cohen:2017a}, the added parameter space of dark matter models
contributes to a larger variety of shapes (e.g., multiple wiggles
during the Cosmic Dawn absorption) for both the global signal and the
r.m.s.\ of the fluctuation. The effect of dark matter on the
absorption trough itself can be very strong, leading to an order of
magnitude increase in the amplitude. Specifically, the deepest
possible absorption in the case with no scattering is $T_{21} = -247$
mK at 120 MHz, while with the scattering the absorption trough can
reach $T_{21} = -2180$ mK at 92 MHz. Additionally, the fluctuations
are enhanced with the peak power shifting from 87.7 mK at 153 MHz to
855 mK at 97 MHz. For the entire ensemble of models without b-DM
scattering the most negative feature of the global signal during
Cosmic Dawn and Reionization ($6<z<35$, $39.5<\nu<202$ MHz) is
anywhere between -247.15 [mK] and -8.02 [mK] and is localized in the
$9.1<z<35$ range ($39.5<\nu<140.6$ MHz); while in all the considered
scenarios with b-DM scattering the maximal absorption is between
-2179.2 [mK] and -2.1 [mK] and can be located anywhere within the
$6<z<35$ range. The strongest fluctuations are expected to have an
r.m.s.\ amplitude between 1.5 and 87.7 mK with the redshift of the
peak power in the range $6.8<z<30$ ($45.8<\nu<182.1$ MHz) in the
models without scattering, and the maximal r.m.s.\ during Cosmic Dawn
($\nu<100$ MHz, $z>13.2$) is 25.3 mK. With scattering the maximal
fluctuation amplitude (due to b-DM scattering only) can be anywhere
between 0 and 855.4 mK at $6.3<z<35$. Finally, with the scattering
affecting the energy budget, gas can be heated either faster or slower
depending on the detailed balance between baryons and the dark
sector. Specifically, we find that the redshift of the heating
transition, i.e., the moment at which the brightness temperature
transitions from absorption to emission during Cosmic Dawn or
Reionization, varies over a wider range when the scattering is
included, and can be anywhere within $6<z_{\rm h}<35$ compared to
$6<z_{\rm h}<30.3$ in the case of no scattering.

Adding the EDGES constraints restricts both the amplitude and the
position of the absorption trough. Namely, the absorption feature of
the compatible models can only be as deep as -1673 mK to -304 mK and
must peak in the narrow redshift range $14.9<z<20.4$ ($66.3<\nu<89.3 $
MHz). This deep absorption trough should be readily accessible to
other global experiments such as SARAS-2 \cite{Singh:2017a,
  Singh:2017b}, LEDA \cite{Price:2017, Bernardi:2016}, and
SCI-HI/PRIZM \cite{Voytek:2014}.  The absorption feature is a
manifestation of the thermal history of the gas and can be used to
constrain the temperature of the gas. To agree with the observations,
the range of $z_{\rm h}$ is restricted to $8.7<z_{\rm h}<17.4$. In other words,
baryons cannot be significantly hotter than the CMB at $z\sim 17$ (the
center of EDGES Low-Band) and should heat up considerably at $z\sim
9-13$. In addition, extremely low heating efficiency is ruled out by
both EDGES High-Band \cite{Monsalve:2018} and SARAS-2
\cite{Singh:2017a, Singh:2017b}. Finally, the scenarios favored by
EDGES yield fluctuations stronger by an order of magnitude than those
previously predicted (assuming collision-less dark matter), with the
peak r.m.s.\ between 8.07 mK and 807.3 mK found at $14.4<z<21.9$
($62.0<\nu<92.2$ MHz). Such strong fluctuations are easily detectable
by interferometric arrays such as HERA \cite{DeBoer:2017} and SKA
\cite{Koopmans:2015}. Note that a strong 21-cm signal is also possible
during the dark ages but it is expected to be at extremely low
frequencies.

Another statistic that can be measured from images taken by 21-cm
interferometers is the probability distribution function (PDF) of the
21-cm brightness temperature (relative to the mean global
temperature). If the function $T_{21}(v_{\rm bdm})$ is monotonic, then
this function can essentially be read off an observed PDF (assuming
that the 21-cm fluctuations are indeed dominated by b-DM scattering),
since the PDF of $v_{\rm bdm}$ itself is known to be Maxwellian
\cite{Tseliakhovich:2010}; however, Fig.~1 shows that in some cases
this function is non-monotonic, so reconstructing it from the 21-cm
PDF will involve model-fitting.

{\it BAO:} The relative b-DM velocity is supersonic at recombination
and, because of the coupling between baryons and photons prior to
recombination, the velocity field bears a strong BAO signature
\cite{Tseliakhovich:2010}. In the absence of b-DM scattering,
$v_{\rm bdm}$ can generate enhanced oscillations in the 21-cm signal by
modulating star formation in primordial halos \cite{Dalal:2010, Visbal:2012}. With
the b-DM scattering, as a result of the dependence of the
cross-section on $v_{\rm bdm}$, the BAO feature in the 21-cm signal is
expected to be even stronger, acting as a smoking gun signature of
b-DM scattering \cite{Barkana:2018}. An enhanced BAO pattern in the
21-cm power spectrum detected by an interferometric array, such as
HERA and SKA, would be a telltale signal of b-DM scattering and
would verify the EDGES detection as well as its dark matter interpretation.

In Fig.~2 we demonstrate the BAO pattern seen in the power spectrum of
the 21-cm power signal from redshift $z=17$ (assuming that DM cooling
dominates and other 21-cm fluctuations can be neglected). To calculate
the power spectrum we generated a distribution of the velocity field
in units of the r.m.s.\ velocity (e.g., see Fig.~1 of
\cite{Barkana:2018}) in a comoving volume of 1.536$^3$ Gpc$^3$. Next,
for each model we transformed the velocity cube to the 21-cm signal
using the $v_{\rm bdm}\rightarrow T_{21}$ mapping from Fig.~1 (top row)
and calculated the power spectrum. The resulting power spectra are
shown in Fig.~2. To highlight the BAO shape (the r.m.s.\ amplitude was
separately shown in Fig.~1) we show the power spectra relative to
their value at $k = 0.1$~Mpc$^{-1}$ and average over 10 independent
realizations of the initial velocity cubes to compensate for the
cosmic variance effect on the largest scales.

\begin{figure}
\begin{center}
\includegraphics[width=3in]{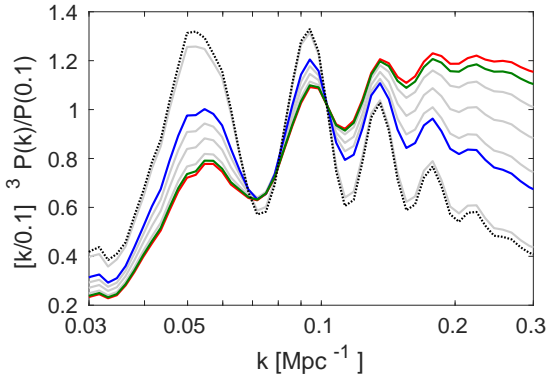}
\caption{Power spectrum of the 21-cm signal versus wavenumber at
  $z=17$. The spectra are shown for the same highlighted models from
  Fig.~1: the model with the strongest absorption (blue), the
  lowest $z_{\rm h}$ (green) and the highest
  r.m.s.\ (red); in addition, several other random models from the
  ensemble compatible with EDGES are shown (grey).  Also shown is the
  power spectrum of $v_{\rm bdm}$ (black dotted curve). The model with the
  strongest BAO has the following parameters: $V_c = 35.5$ [km
    s$^{-1}$], $f_*=0.3$\%, $f_X = 0.0721$, $\alpha = -1.5$,
  $\nu_{min}=0.4$ [keV], $\tau = 0.071$, $R_{\rm mfp}=30$ [Mpc],
  $m_\chi=0.56$ [GeV] and $\sigma_1 = 4.6\times 10^{-20}$ [cm$^2$].
  In order to highlight the BAO shape, all the curves are normalized
  to unity at $k = 0.1$ [Mpc$^{-1}$]. }
\end{center}
\end{figure}

{\it Conclusions:} The recent detection by EDGES Low-Band, if indeed
cosmological, requires a new theoretical explanation beyond the
standard astrophysical model. In this Letter, considering b-DM
scattering as a viable mechanism to produce the observed absorption,
we have surveyed the resulting parameter space of astrophysical and
dark matter properties. We have shown that the expected global signal
and r.m.s.\ of the fluctuations vary over a much larger range than
before. The velocity-dependent cross-section imprints enhanced BAOs
which could be used to constrain dark matter theories.

\bibliography{SCINC}{}

\end{document}